# Prediction of non-identical particle correlations for the Beam Energy Scan program

Katarzyna M. Poniatowska[1a]

[1]Warsaw University of Technology, Koszykowa 75, 00-662 Warsaw, Poland

**Abstract.** Femtoscopy of two non-identical particles in heavy ion collisions enables one to study the space-time asymmetry in the particle's emission process. Theoretical studies based on EPOS model performed for collision energies from the Beam Energy Scan program in STAR allow us to investigate the dependence of source sizes and dynamics effects. Obtained information will enable us to predict the collective behaviour of femtoscopic particle's source.

## 1 Introduction

During the heavy ion collisions the particle's emission sources are created, which are very small and live shortly. The order of magnitude of source size is $10^{-15}m$ and the lifetime is $10^{-23}s$, they can not be measured directly. It is possible to use the similar method as Hanbury Brown and Twiss (HBT) used in astronomy - the two-particles correlation femtoscopy[1]. After measuring momentum of two particles, calculating the difference between them, and repeating this steps for many pairs, the correlation function C(q) can be calculated as a function of momentum difference (q = $p_1$ - $p_2$ ; p – particle momentum). The size of the source which emitted those pairs can be extract.
If the correlation function for non-identical particles e.g. pions and kaons is calculated, the space-time asymmetry in emission process can be found[2]. Basing on the correlation function of nonidentical particles. Information deduced from measured asymmetry in emission process corresponds to particles emission sequence: particle of which type is statistically emitted closer/further from source center and/or which are emitted first/second. STAR experiment measured non-identical particles correlation for $\sqrt{s_{NN}}=130 GeV$, it was proved that statistically pions are emitter closer to the center of the source or/and later, than kaons [3]. Investigation of these effects are the main motivation for the study of non-identical particle correlations femtoscopy.

The main task of the Beam Energy Scan (BES) program at the Solenoid Tracker at RHIC (STAR)[4] is to scan the QCD phase diagram with heavy ion collisions at $\sqrt{s_{NN}}=(7.7-62.4)GeV$ to find information about the $1^{st}$ order phase transition between Hadron Gas and Quark Gluon Plasma which is located between $1^{st}$ order phase transition and "cross-over" transition.
Information derived from studies of non-identical particle combination correspond to source size and space-time asymmetry in the emission of two types of considered particles. Analysis of two-particle correlations for a variety of collision energy (e.g. for BES program) enables to find possible relation of (source size, space-time asymmetry) and collision energy. To predict such relations data from EPOS [5] model were used.

[a]Corresponding author: poniatowska@if.pw.edu.pl

EPOS is a parton model, with many binary parton-parton interactions, where each one creates a parton ladder. This model contains few dependences[5]: Energy-sharing (for cross section calculation and particle production); Parton Multiple scattering; Outshell remnants; Screening and shadowing via unitarization; Splitting; Collective effects for dense systems ( LHC energies ).

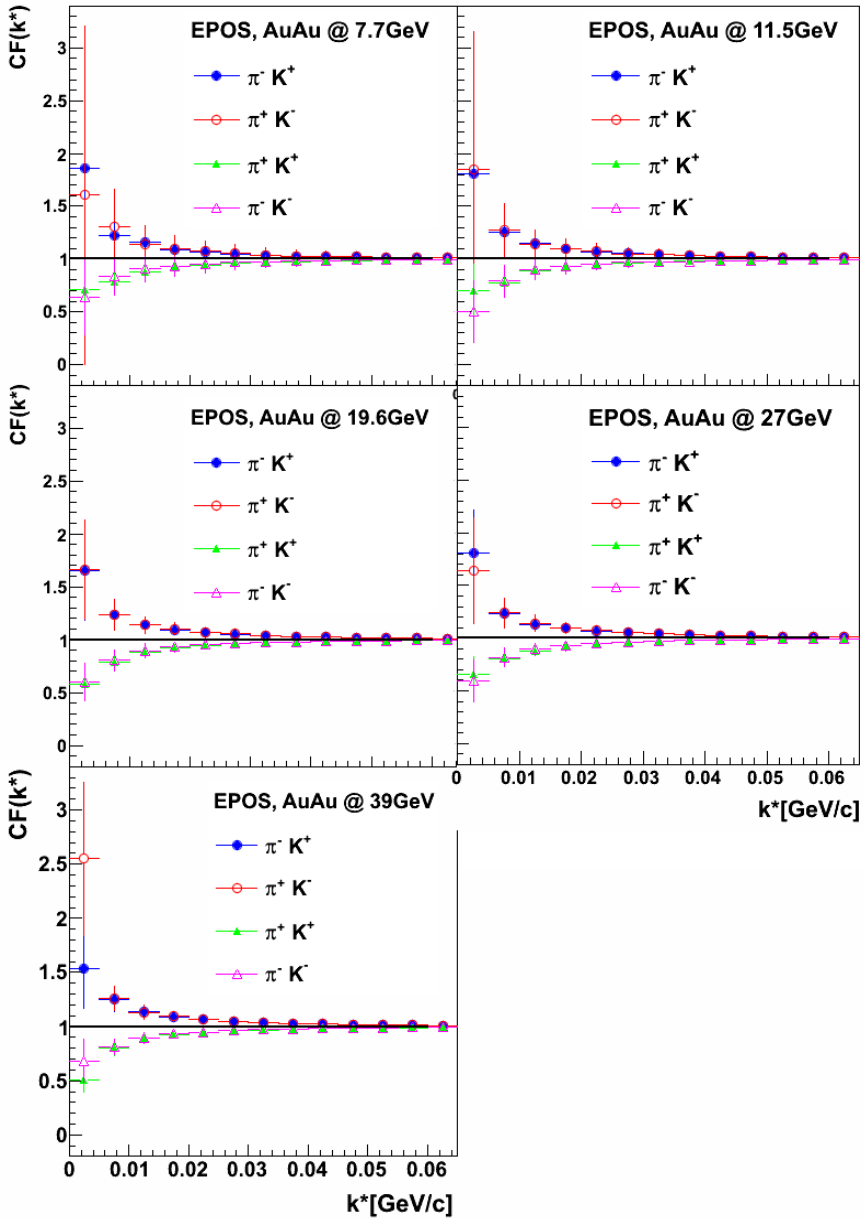

**Figure 1.** Correlation functions for all charged pion-kaon systems for collision energies $\sqrt{s_{NN}}=(7.7,11.5,19.6,27,39)GeV$, for data from EPOS model.

## 2 Correlation femtoscopy

The correlation functions of two particles emitted with small relative velocities are calculated in the momentum difference variable k*. If first particle has momentum $p_1$ and energy $E_1$, and second particle has momentum $p_2$ and energy $E_2$ then $k^* = \frac{1}{2}\sqrt{(p_1-p_2)^2 - (E_1-E_2)^2}$. The correlation function is defined as a ratio: C(k*) = A(k*)/B(k*), where A(k*) is the signal and B(k*) is the background. Pairs of correlated particles which came from the same collision entered into the numerator A(k*) and the pairs of uncorrelated particles, from different events, into the denominator B(k*). The shape of correlation function depends on FSI interactions. The dominating FSI interaction

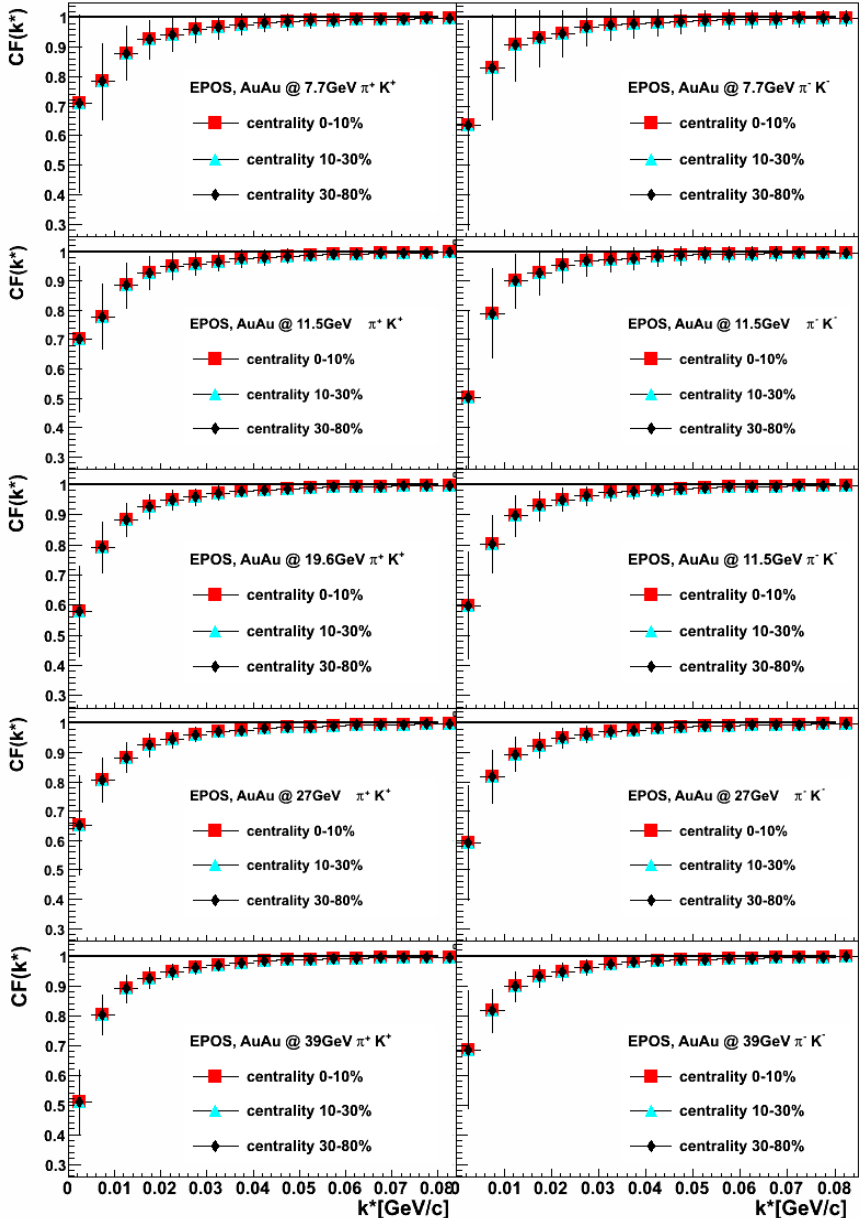

**Figure 2.** Centrality dependence for correlation functions for the systems with the same signs for $\sqrt{s_{NN}} = 39 GeV$.

for non-identical particles such as pion-kaon correlation femtoscopy is the Coulomb force[6].

## 2.1 Source size

The two-particles emission source size can be extracted using the femtoscopy method. First the source parametrization has to be assumed e.g. the source density should be considered as a gaussian[6], to fit the correlation function the CorFit [7] tool is use. Is expected that the source size derived from two-particle correlation function with increases with the energy of collision. The STAR experiment measured the size of the π – K emission source as *(12.5 ± 0.4)fm*, for Au+Au at $\sqrt{s_{NN}}=130 GeV$ [3]. It was expected, that for lower collision energies sources should be smaller assuming gaussian source distribution due to e.g. lower particles multiplicities [8].

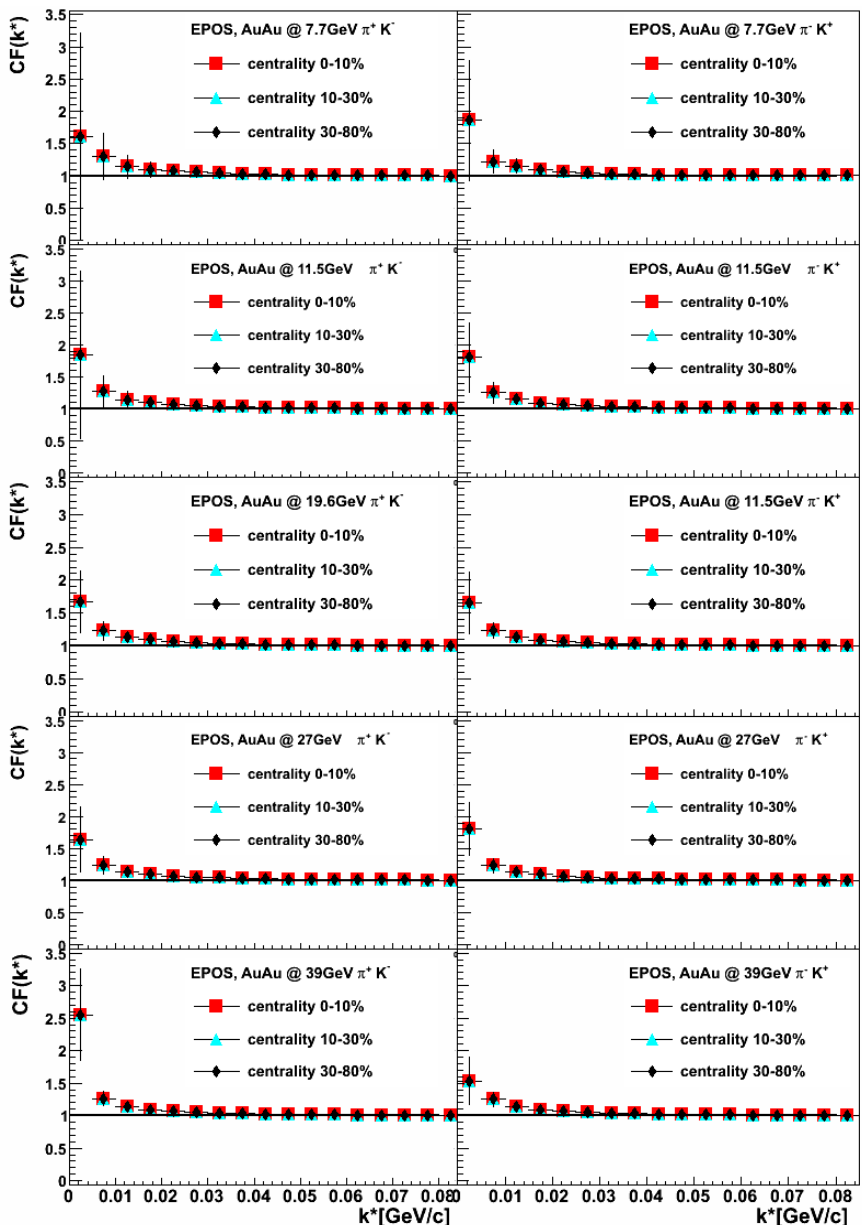

**Figure 3.** Centrality dependence for correlation functions for the systems with the different signs for $\sqrt{s_{NN}} = 39 GeV$.

## 3 Results

In this analysis to calculate π – K correlation functions $10^5$ events with Minimum Bias Au+Au collisions are used for following collisions energies $\sqrt{s_{NN}}=(7.7, 11.5, 19.6, 27, 39) GeV$.

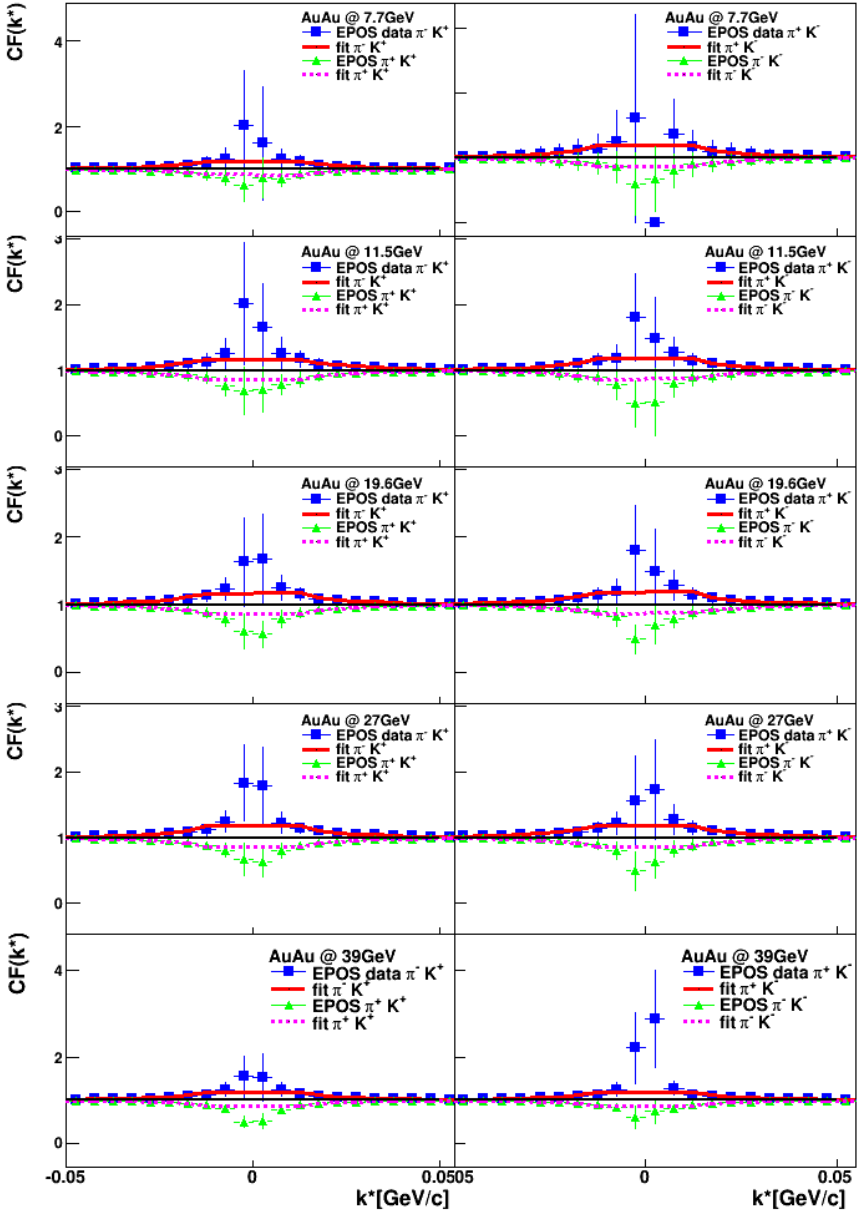

**Figure 4.** Theoretical fits for correlation functions of all pion-kaon systems for all collision energies $\sqrt{s_{NN}}=(7.7, 11.5, 19.6, 27, 39) GeV$, for data taken from EPOS model.

All systems are analysed for different collision centralities (*0-10%* - most central, *10-30%* - medium central, *30-80%* - peripheral).

### 3.1 Correlation functions

Two-particle correlation functions are presented in Fig. 1. The correlation function for identical signs combination indicated anti-correlation due to depletion of correlated pair being repelled by Coulomb force. The correlation function for the opposite charges introduce correlation due increased number of pair being a result of Coulomb attraction. We did not observed a relation between correlation strength and the collision energy.

### 3.2 Centrality dependence

In order to accurately examine the correlation depending on the type of collision, a separate analyses performed for different centralities. The results of these analyses are presented in the Fig. 2, 3. The shape of the correlation functions do not change with the centrality of the collision. This relation is observed for all five systems energy of particles of the same sign Fig. 2 and different signs Fig. 3. The fact that data do not depend on collision centrality may be caused by the leak of hydrodynamic systems evolution in the EPOS model.

### 3.3 Source sizes

In order to determine the source sizes the gaussian source distribution are assumed, using such parametrization the source size are deduced for four different two-particles systems for Minimum Bias data. The theoretical fits are done for data correlation functions Fig. 4. Designated sizes of particle emission sources as a function of collision energy are summarized in the Fig. 5. As expected, the sizes of the investigated sources are smaller than the sizes of the source measured in the STAR experiment for $\sqrt{s_{NN}}=130 GeV$. In the EPOS model, the size of the emission sources are similar for all sign systems. We do not observe dependence of the size of the source on the collision energy. This may be caused by the leak of hydrodynamical description of system evaluation in EPOS model.

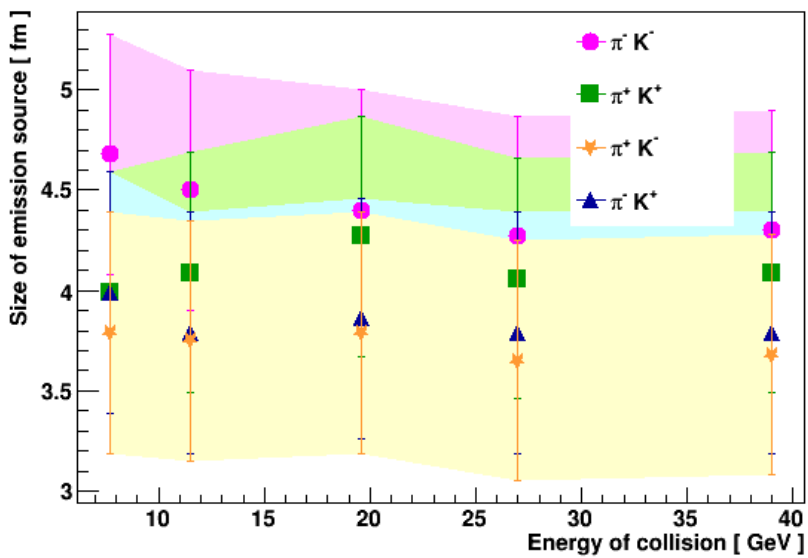

**Figure 5.** Sizes of emission sources in function of collision energy (in the center of the mass) for all pion-kaon sign systems for data from EPOS model.

## 4 Summary

Analysis are conducted of data from EPOS model for Au+Au collisions with energies $\sqrt{s_{NN}}=(7.7, 11.5, 19.6, 27, 39)GeV$. Femtoscopy correlation functions of non-identical particles pion-kaon for four different sign combinations are computed. The shape of the correlation function is described by Coulomb interaction.

We do not registered significant differences in both shape and source size from the correlation function. There is no big difference in EPOS model between correlation functions of the same sign particles systems ($\pi^- K^-$ and $\pi^+ K^+$) and for correlation functions for different sign particles systems ($\pi^- K^+$ and $\pi^+ K^-$). Association between the strength of interaction and collision centrality for $\sqrt{s_{NN}}=(7.7, 11.5, 19.6, 27, 39)GeV$ are not observed. Source sizes calculated for BES program are smaller then size from collision energy $\sqrt{s_{NN}}=130GeV$, but we do not observed energy dependence in source sizes in EPOS model. To compare the STAR results from BES with the model, we need the model with hydrodynamics.

## Acknowledgements

This work was supported by the Grant of National Science Centre, Poland, No: 2012/07/D/ST2/02123